\documentclass[epjST,useAMS,aas_macros]{svjour}
\usepackage{graphics}
\usepackage{subfig}
\usepackage{tabularx}
\usepackage{epsfig}
\usepackage{multicol}
\usepackage{pstricks,pst-grad}
\usepackage{tablefootnote}
\usepackage{hyperref}

\newcommand{\aap}{    {\it Astron. Astrophys.}}

\newcommand{\ao}{     {\it Appl. Opt.}}

\newcommand{\apjs}{   {\it Astrophys. J. Suppl.}}
\newcommand{\apjl}{   {\it Astrophys. J. Let.}}

\begin{document}
\title{Membership of Stars in Open Clusters using Random Forest  with Gaia Data}
\author{Md Mahmudunnobe\inst{1} \and Priya Hasan\inst{2}\fnmsep\thanks{\email{priya.hasan@gmail.com}} \and  Mudasir Raja\inst{2} \and S N Hasan\inst{2} }

\institute{ Minerva Schools at KGI, San Francisco, California 94103, USA. \and Maulana Azad National Urdu University, Gachibowli, Hyderabad  500 032, India.}
\abstract{
 Membership of stars in open clusters is one of the most crucial parameters in studies of star clusters. Gaia opened a new window in the estimation of membership because of its unprecedented 6-D data.  In the present study, we used published membership data of nine open star clusters  as a training set to find new members from Gaia DR2 data using a supervised random forest model with a precision of around 90\%. The number of new members found is often double the published number. Membership probability of a larger sample of stars in clusters is a major benefit in determination of cluster parameters like distance, extinction and mass functions. We also found members in the outer regions of the cluster and found sub-structures in the clusters studied.  The color magnitude diagrams are more populated and enriched by the addition of new members making their study more promising. 
} 
\maketitle

\section{Introduction}
\label{intro}
Star clusters are the building blocks of galaxies and are the key to understanding the formation and evolution of stars and galaxies  \cite{lynga82} \cite{janes94} \cite{khar05} \cite{friel95}  \cite{bon06}.  
Star clusters are seen as an over-density and are a sample of stars in the same region of the sky,  gravitationally bound and formed from the same molecular cloud. Hence the  member stars are approximately at the same distance, of the same age and only differ in mass. These stars move with a common velocity, which is an imprint of their formation process.
The well-known open cluster catalogues are \cite{2002A&A...389..871D} and \cite{2013A&A...558A..53K} and they list about 3000 open clusters of the Milky Way.

In the study of star clusters, one of the most crucial parameters is cluster membership. To identify members, we look for stars in the same region of the sky, at the same distance and with a common velocity. However, such samples may be contaminated as there can be field stars with similar velocities or at similar distances. Also,  photometric and spectroscopic (kinematic) data of cluster stars are generally not available for all the stars in a cluster. For most stars, we determine  a  membership probability,  which  is  a  non-trivial  problem.  In the presence of only photometric data, the standard method used is a photometric envelope in the color magnitude diagram (CMD) due to which variable stars, outliers, etc. are difficult or impossible to be identified as members. Spectroscopic data is generally available only for a few selected bright stars of the cluster putting limitations on membership determination. Various methods have been used for membership determination using stellar positions, proper motions, parallaxes,
radial velocities, photometry and their combinations \cite{1958AJ.....63..387V,1977A&AS...27...89S,1990A&A...237...54Z,2020MNRAS.492.5811B}.

The second Gaia data release DR2  \cite{2016AA...595A...1G}\cite{2018AA...616A...1G} (and references therein) contains precise astrometry at the sub-milliarcsecond level and homogeneous three-band photometry for about 1.3 billion sources. This can be used to characterize a large number of clusters over the entire sky. Using Gaia data DR2, \cite{2018AA...618A..93C} provided a list of 1229 clusters with membership data and derived parameters, in particular, mean distances and proper motions. They applied an unsupervised membership assignment code, UPMASK \cite{2014A&A...561A..57K}, which is primarily based on the $K$-means clustering algorithm to detect the cluster members and then using random sampling to assign the membership probability. This work was extended by the same authors for a  final list of members for 1481 clusters \cite{refId0}.

In this paper, we used  membership data from \cite{2018AA...618A..93C} as our training data to find  new  members in a sample of nine open star clusters. We discuss the results obtained and present the advantages of applying such a technique in membership determination of star clusters.

The paper is structured as follows: Section~1 is the Introduction, Section~2 describes the Gaia DR2 Data and Sample Selection, Section~3 has  the details of the Machine Learning technique applied in the problem, Section~4 presents the Results and in Section~5 we discuss our Conclusions.

\section{Data and Sample Selection}
\label{sec:1}

\begin{figure}[h]
\includegraphics{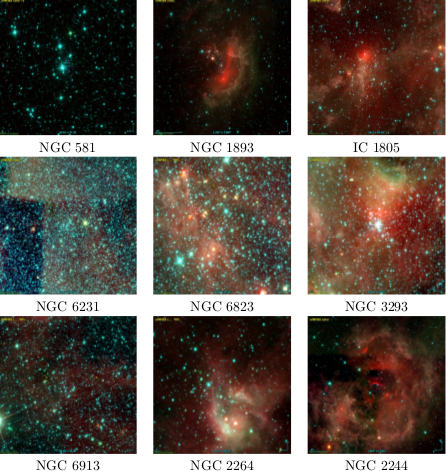} 
\caption{WISE images of the cluster sample: NGC 581, NGC 1893, IC 1805, NGC 6231, NGC 6823, NGC 3293, NGC 6913, NGC 2264, NGC 2244. In all images, North is up, East is left.}
\label{clusfig}       
\end{figure}

We selected a sample of nine young clusters {\it viz.} NGC 581, NGC 1893, IC 1805, NGC 6231, NGC 6823, NGC 3293, NGC 6913, NGC 2264 and NGC 2244 (Fig. \ref{clusfig}).  The sample has  clusters with ages ranging from 1.3--20 Myr, at galactocentric distance $R_{GC}$ ranging from 7.3--14.5 kpc and at varying  galactic latitudes~$l$ and  longitudes~$b$ which have been listed in Table~\ref{clusterdata}. This sample will also be useful to study the effectiveness of this procedure as the clusters vary in their position in the galaxy and their individual structures and parameters. 

\begin{table*}[h]
\small

\caption{Basic cluster parameters}
\label{clusterdata}
\begin{tabular}{llllllll}
\hline

Cluster&   $l$ &$b$ & Ang.Dia & Distance &  $E(B-V)$ & log(age) & $R_{GC}$   \\
         & deg     & deg    &arc min & pc         & mag  &log(yr) & kpc\\ \hline
NGC 581  & 128.05  & -01.80 & 5.0    & 2194  & 0.38 & 7.3    &10.0 \\
NGC 1893 & 173.59  & -1.68  & 25     & 6000 & 0.45 & 6.5    & 14.5\\ 
IC 1805  & 134.73  &  0.92  & 20     & 2344 & 0.87 & 6.1    & 10.3 \\
NGC 6231 & 343.46  & +01.18 & 14.0   & 1243 & 0.85 & 6.5    & 7.4\\
NGC 3293 & 285.86  & +00.07 & 6.0    & 2327 & 0.26 & 7.0    & 8.2\\
NGC 6913 & 76.91   & +00.59 &10.0    & 1148 & 0.74 & 7.1    & 8.3\\ 
NGC 2264 & 202.94  &  +02.2 &39.0    &  667 & 0.05 & 6.9    & 9.1\\  
NGC 2244 & 206.31  & -02.07 &29.0    &  1445& 0.46 & 6.9    & 9.8\\ 

\hline
\end{tabular}
\end{table*}

%

\section{Machine Learning and the RF Method}
Machine Learning (ML) has been used very effectively in various astrophysical problems, including the problem of membership in star clusters. 
The Random Forest (RF) method was used to identify reliable members of the old (4 Gyr) open cluster M67 based on the high-precision astrometry and photometry taken from the Gaia DR2 and was  found to be very effective for membership determination of open clusters \cite{Gao_2018}. Once cluster members are determined, cluster parameters like  distance, proper motion, radial velocity and spatial structure of M67 were found. The method found not only likely cluster members but also possible escaped members, which may lie outside the tidal radius of the cluster.

Membership of stars in  NGC 188 was studied using a machine-learning-based method which  combined two widely used algorithms: spectral clustering (SC) and random forest (RF) \cite{2014RAA....14..159G}. The $K$-means clustering algorithm was used to study  membership of NGC~188 and NGC~2266 \cite{2016ExA....42...49E}. 

Reliable memberships and fundamental astrophysical properties of the clusters NGC~6791 and  NGC~6405 were found using the astrometric and photometric data of the Gaia DR2 \cite{2020Ap&SS.365...24G} \cite{2018AJ....156..121G}. ML based membership
determination algorithms for open clusters in the absence of a priori information of cluster parameters was done by \cite{2021MNRAS.502.2582A} using k-nearest neighbour ($kNN$) and Gaussian Mixture Model ($GMM$).

 RF is an ensemble machine-learning technique mainly used for classification and regression tasks. In recent years, RF has been widely used to investigate a variety of astronomical problems. RF
can easily handle high-dimensional data without data normalization  and obtain class membership
probabilities that allow us to isolate the most likely cluster members. RF has very short execution time  which allows us to effectively handle large,
high-dimensional datasets. However, as a supervised ML technique, the effectiveness of RF
strongly depends on a reliable training set; a poor training set will yield unreliable results and hence is a major caveat.
\label{sec:2}

\subsection{Training Data}
 In order to train our model, we need samples of both member stars and non-member stars. For each of the clusters, we take all the members provided by \cite{2018AA...618A..93C} as our training set.  The above authors also  provided a membership probability for each of the member stars.
 In their method, any star with a membership probability $> 0 $ is a possible member. 
 
We  added additional criteria to filter bad data, and only included stars whose parameters from the Gaia DR2  satisfied $parallax\_over\_error > 3$,  $pmra\_error < 0.3$ and $pmdec\_error < 0.3$\footnote{Right Ascension (RA) is denoted by $\alpha$ and declination (Dec) by $\delta$. Proper motion in RA is pmra ($\mu_{\alpha}$) and  proper motion in declination is pmdec ($\mu_{\delta}$) and both are in units of milli-arcseconds per year (mas/yr). Parallax is denoted by $\omega$ and has units of (mas).}. Henceforth we will refer to these member stars of the training set as CG members and we reported the total CG members in each clusters in Table \ref{memdata}.

The radius of open clusters is a difficult parameter to define. The primary reason behind this is the lack of precise data to identify members. Earlier observations detected only a few members closer to the cluster center, thus reporting a smaller radius. Improved observations lead to detection of members in outer regions, thus deriving a larger radius of clusters. For example, \cite{2018AA...618A..93C} reported a larger radius for many clusters using the precise Gaia data.  In order to find new members in outer regions, we used a search radius around each cluster which was double the maximum radius found in CG members, where the maximum radius is defined as the distance of the farthest members from the center. The only exception in this case was NGC 2264, where the presence of an additional sub-clump required us to increase our search radius to thrice the cluster radius. 

For the sample of non-members, we took stars from a concentric ring  outside the search radius as shown in Table \ref{memdata}. To balance our training data and avoid any over-fitting, we took a random sample of an equal number of non-members and members for the cluster. Thus the total number of stars in the training set for each cluster is twice the CG members of the cluster.

Similar to \cite{2018AA...618A..93C}, we used $\mu_{\alpha}$, $\mu_{\delta}$ and $\omega$ to build our model since all the members of a cluster should be concentrated in a small region in the 3D space of $\mu_{\alpha}$, $\mu_{\delta}$ and $\omega$, while  the non-members should lie outside  this region. For an open cluster, of course, the members lie in the same region of the sky ($\alpha$, $\delta$). But we do not  use $\alpha$, $\delta$ as training features as we have already selected our search radius around the cluster center. We also did not use the distance from the cluster center as a feature because the non-members are chosen from an outer ring area and any supervised ML method will lead to an over-fitting to get a higher accuracy by setting a cutoff with this distance parameter.

\begin{figure}[h]
   \centering
    \includegraphics[width = 0.8\linewidth]{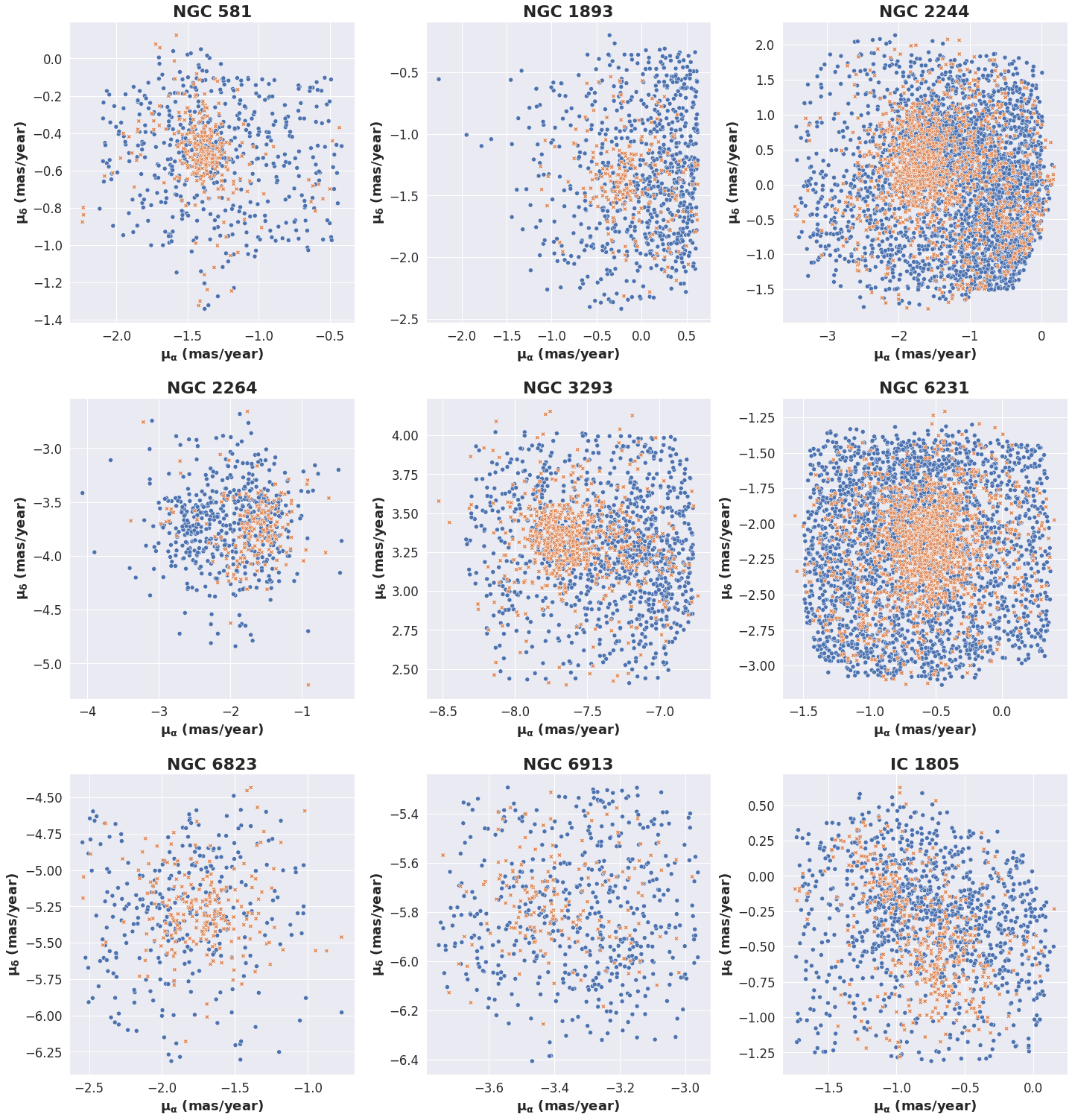}
    \caption{Proper motions  of the CG (orange) and new (blue) members for our sample of clusters. }
    \label{pmall}
\end{figure}

\begin{figure}[h]
    \centering
    \includegraphics[width = 0.8\linewidth]{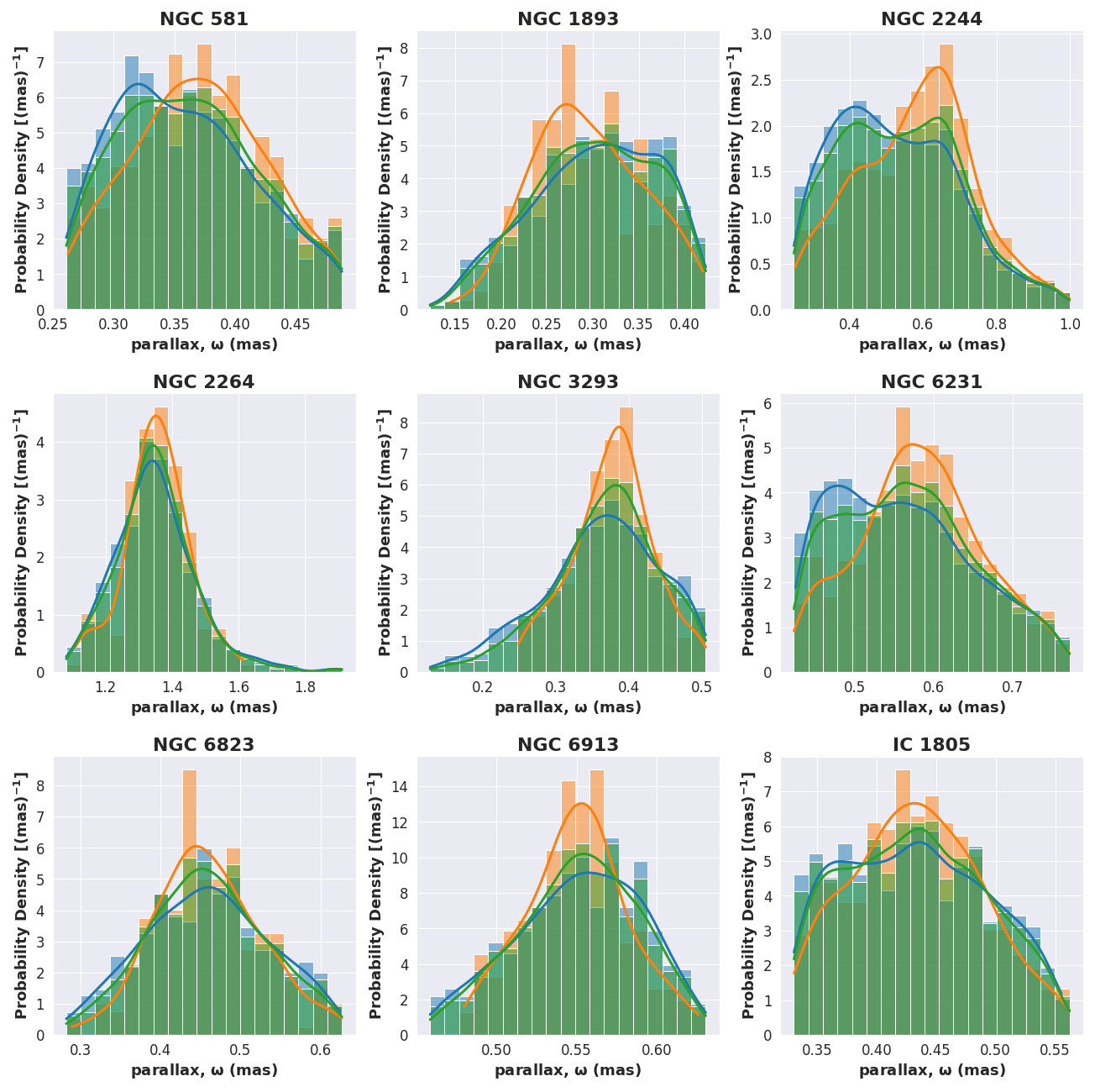}
    \caption{Parallax distribution of CG (orange), new (blue) and combined (green) members for all nine clusters which  shows that the new members are very similar in parallax to the CG members.}
    \label{parall}
\end{figure}

\begin{figure}[h]
    \centering
    \includegraphics[width = 0.8\linewidth]{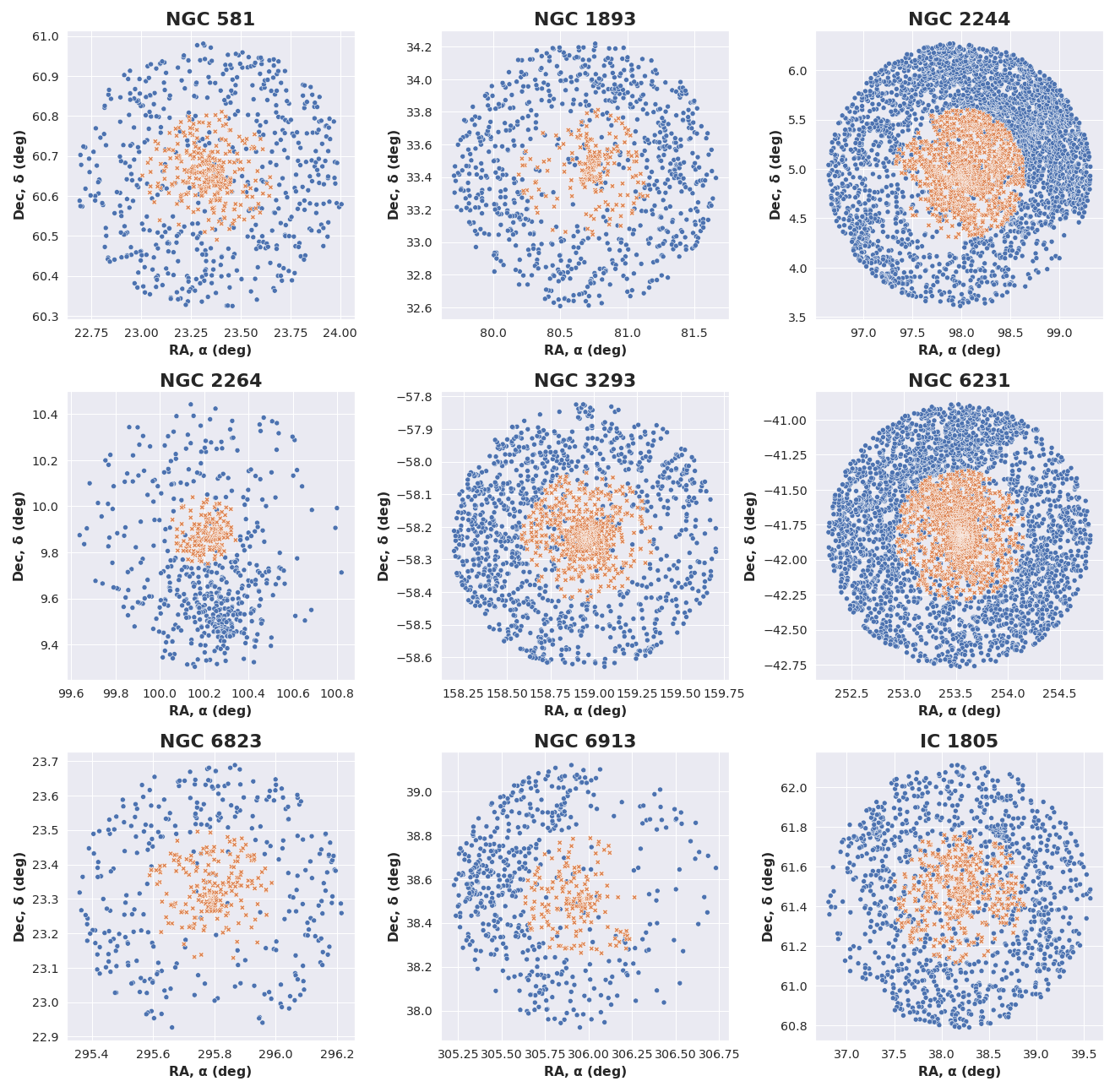}
    \caption{Distribution of the CG members (orange) and new members (blue) in sky coordinates, showing a large number of identified new members on the boundary of the clusters.}
    \label{sky}
\end{figure}

\begin{figure}[h]
    \centering
    \includegraphics[width = 0.8\linewidth]{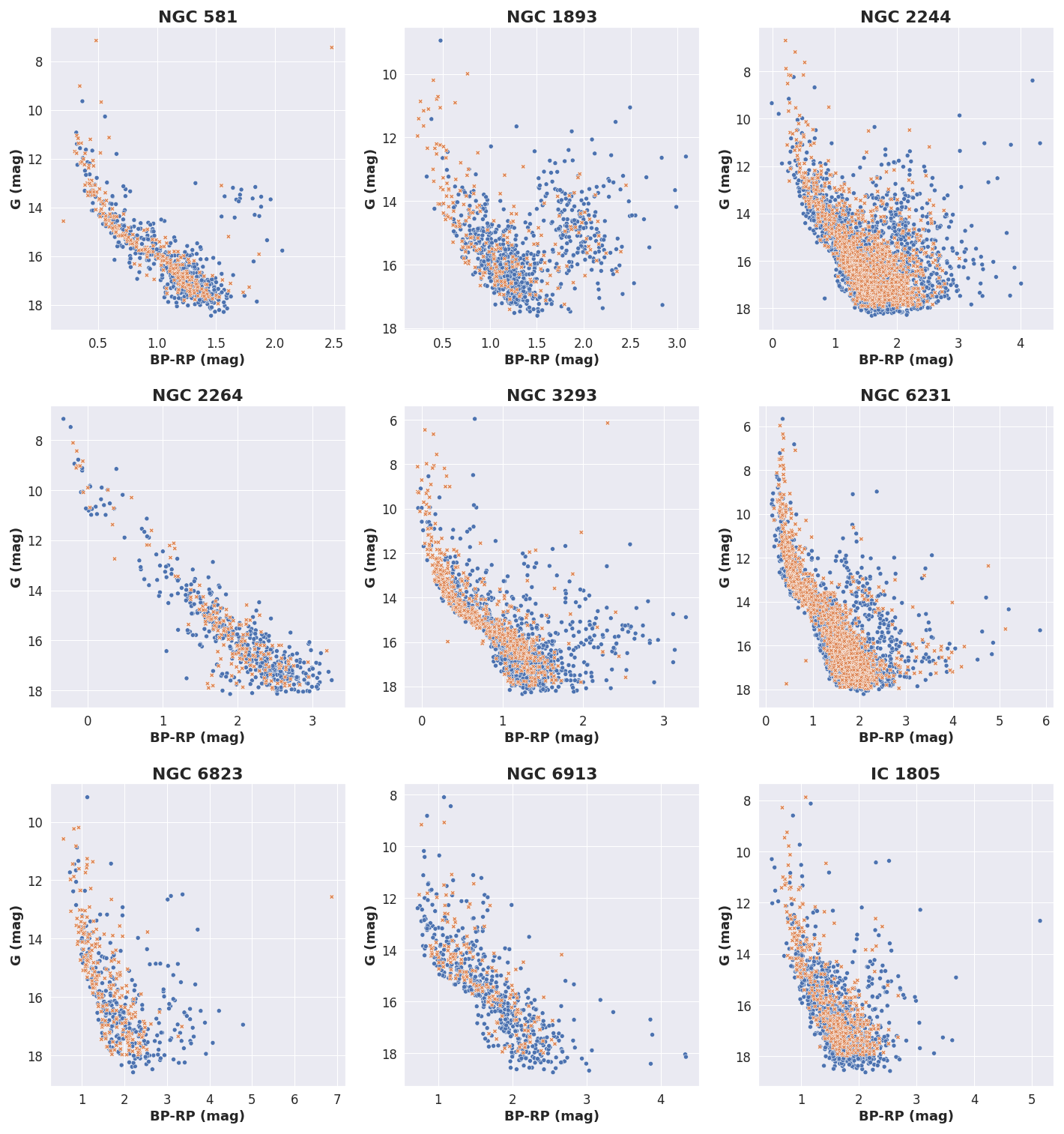}
    \caption{Color-Magnitude Diagram of CG (orange) and new members (blue) for all nine clusters. We can see that this method is effective in deriving variable stars, pre-main sequence stars and non-main sequence  stars and enriches the CMD.}
    \label{cmall}
\end{figure}

\begin{figure}[h]
    \centering
    \includegraphics[width=0.82\linewidth]{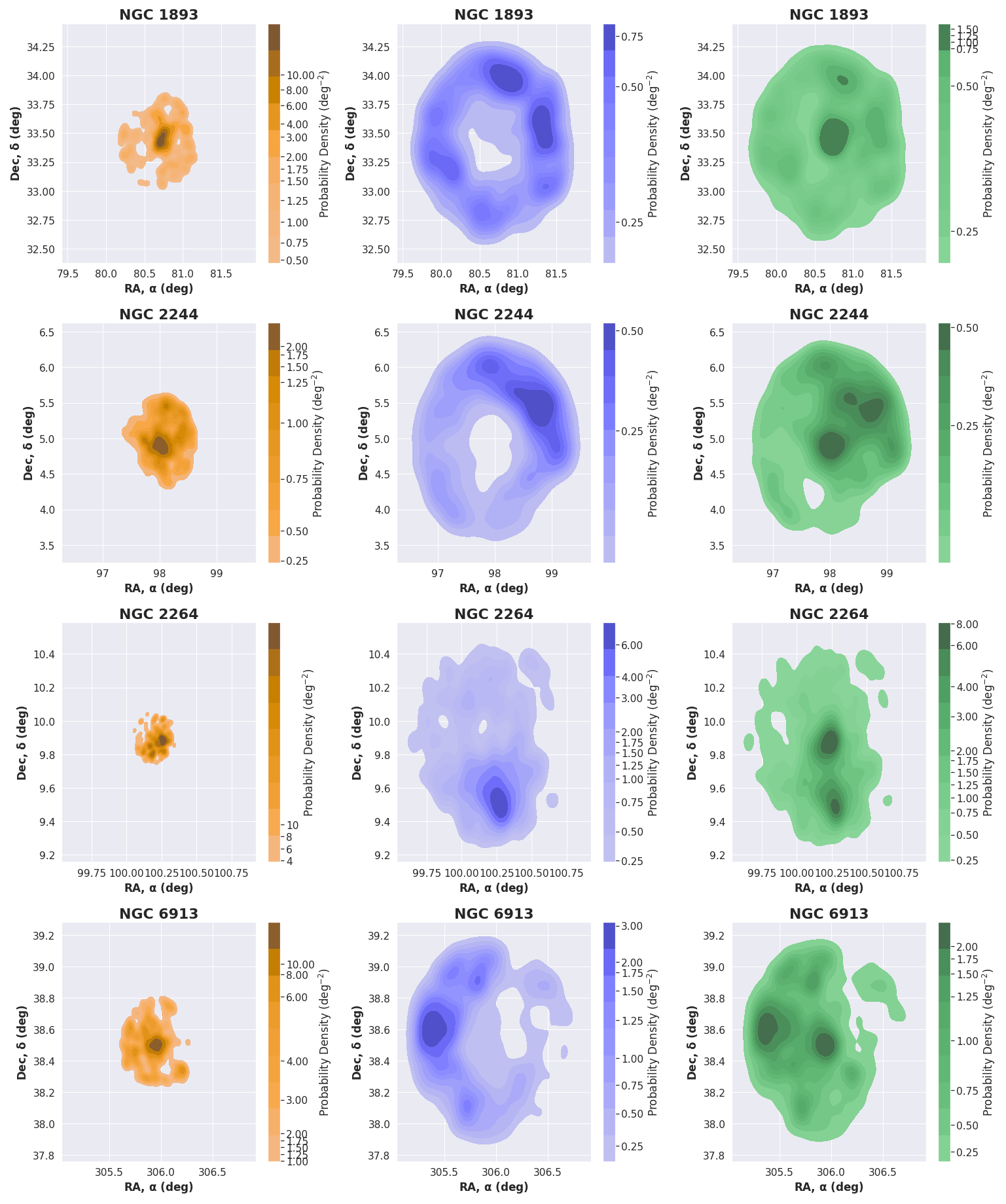}
    \caption{Sky plot of CG (orange),  new members (blue) and combined (green) members of the sample clusters with the probability density of the stars. Note: NGC 1893, 2244, 2264 and 6913 show sub-cluster regions.}
    \label{skyall1}
\end{figure}

\begin{figure}[h]
    \centering
    \includegraphics[width=0.8\linewidth]{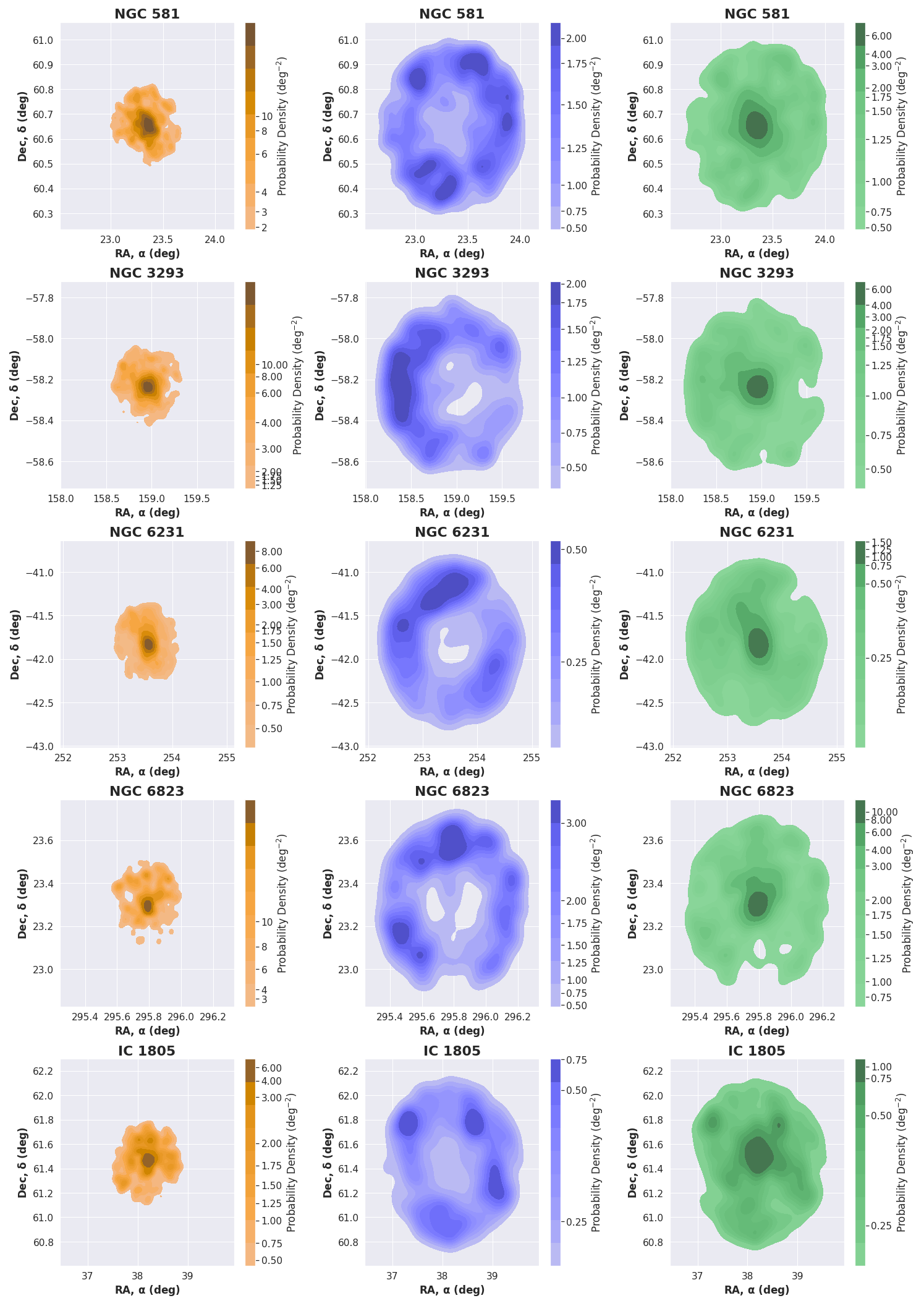}
    \caption{Sky plot of CG (orange),  new members (blue) and combined (green) members of the sample clusters (contd)}
    \label{skyall2}
\end{figure}

\begin{figure}[h]
    \centering
    \includegraphics[width = 0.8\linewidth]{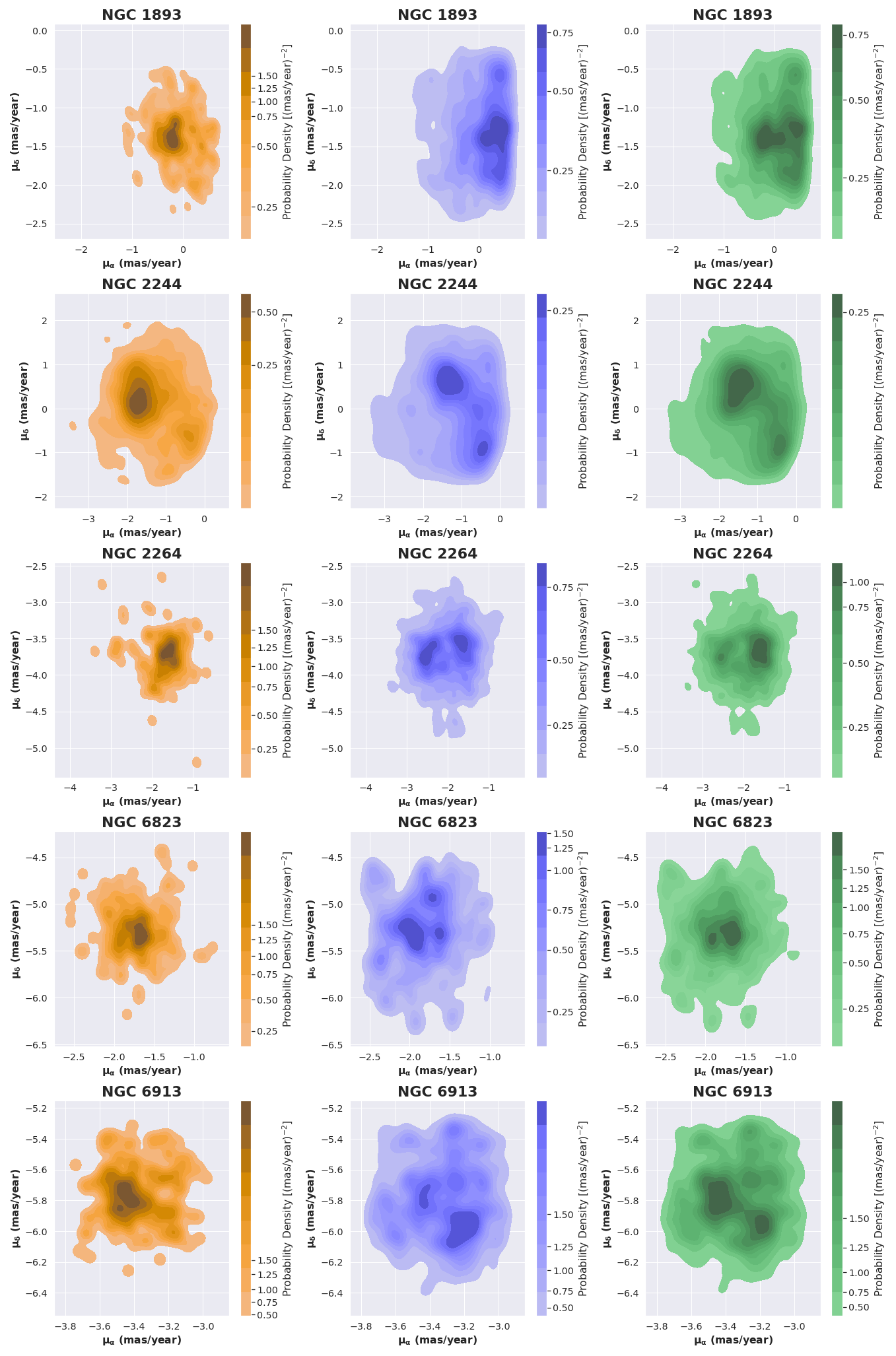}
    \caption{Proper Motion Plot of CG (orange),  new members (blue) and combined (green) members of the sample clusters with the probability density of the stars. These clusters show smaller additional clumps in the proper motion space.}
    \label{pmall1}
\end{figure}

\begin{figure}
    \centering
    \includegraphics[width = 0.8\linewidth]{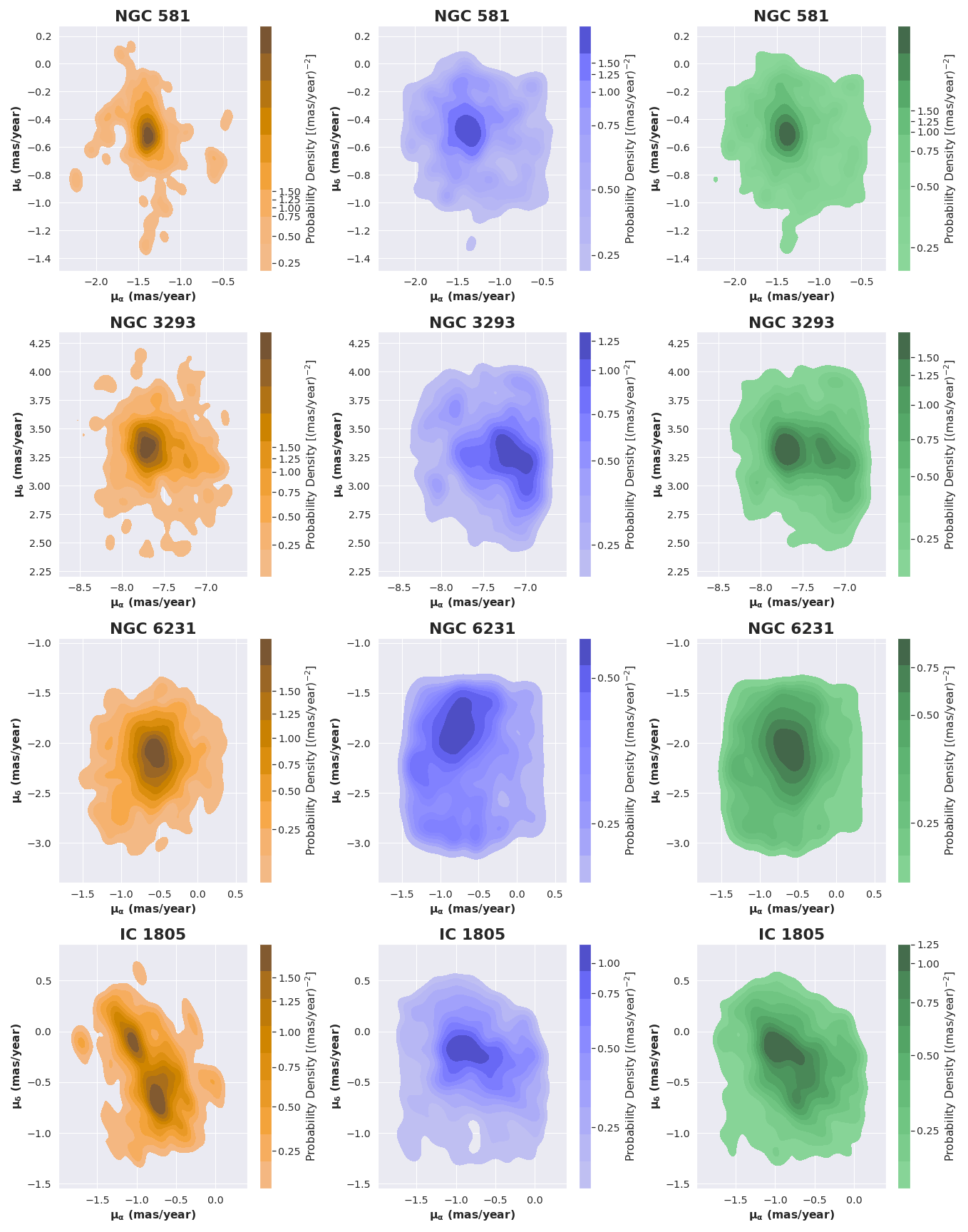}
    \caption{Proper Motion Plot of CG (orange),  new members (blue) and combined (green) members of the sample clusters (contd)}
    \label{pmall2}
\end{figure}

\begin{figure}
    \centering
    \includegraphics[width = 0.8\linewidth]{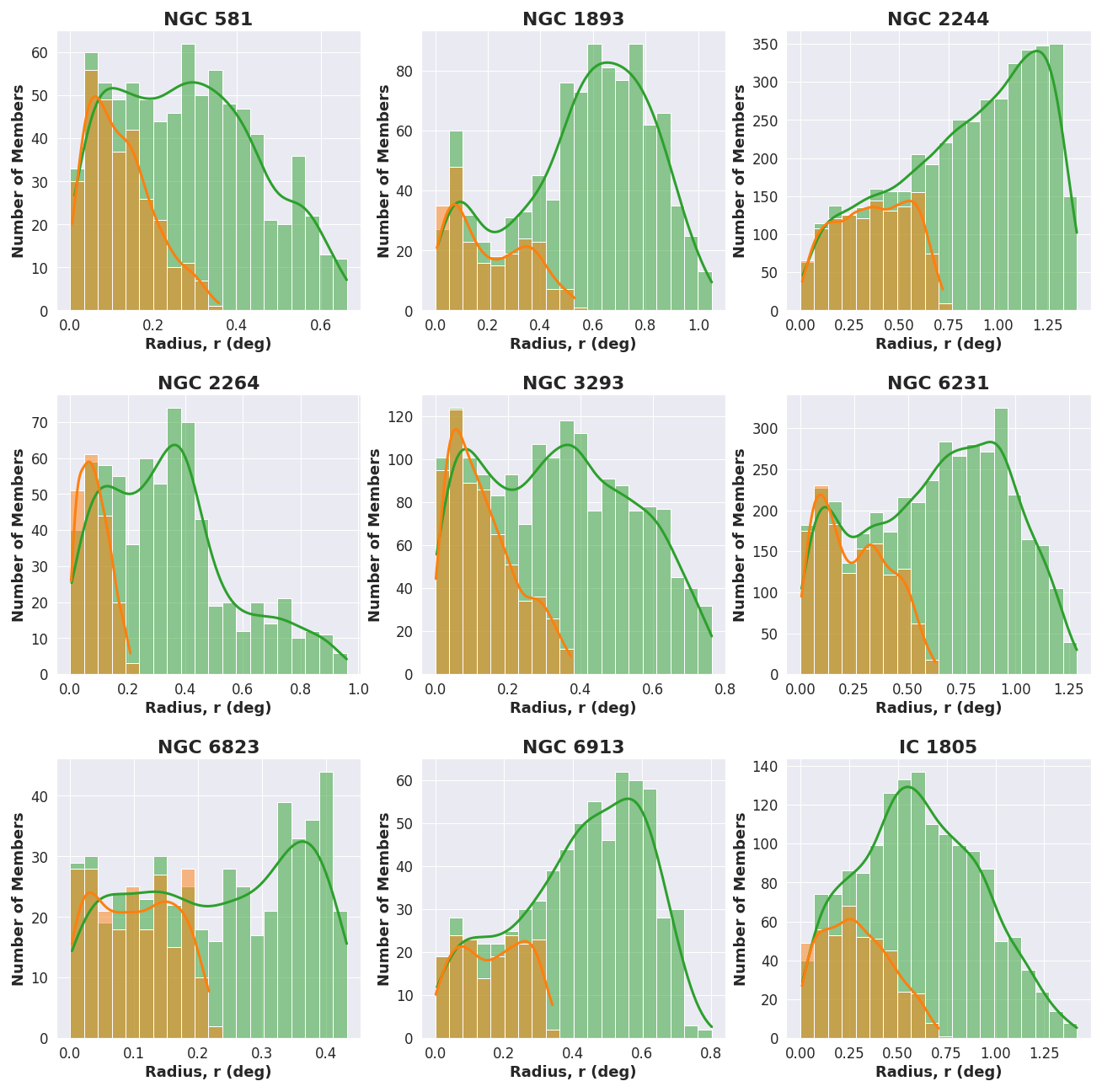}
    \caption{Stellar distribution of the clusters with CG (orange) and combined (green) members of the sample. }
    \label{sd1}
\end{figure}


\subsection{Evaluation Metric}
The members of open clusters can be used to study the initial mass function, stellar evolution and dynamics and the star formation process in clusters. Having a large number of members helps to reduce errors in determination of parameters, while wrongly identified members can lead to misleading results. Thus it is a common practice to apply strict criteria to avoid any non-members being selected as members. As we would like to avoid false positives as cluster members, we will use precision as our metric to evaluate the performance of the model, which is defined as  $$\frac{\rm True\ Positive}{\rm True\  Positive+False \ Positive}$$
We will use this to hypertune the model parameters.

\subsection{Hyper-tuning Model Parameters}
Cluster members occupy a closely bound region in the 3D space of ($\omega,\mu_{\alpha},\mu_{\delta}$) in a spherical or normal distribution, while non-members are randomly distributed outside this region. If we analyze and compare the possible supervised classification models for this problem, then firstly we can say that any linear parametric model (i.e. linear support vector machine (SVM)) would not be  suitable as the separation is not linear but spherical. The other possible options  we could explore are SVM with a suitable kernel, Naive Bayes with an Gaussian distribution and $kNN$ after scaling the features to reduce over-fitting. Another suitable model is the decision tree method as we have smaller number of features and the members are quite separated. But using a single decision tree can increase the over-fitting which results in a less stable model. Therefore in this paper we explore with RF method which can decrease the bias and increase the variance by using a large number of trees. 

\begin{table}[h!]
\small
\caption{The random selection grid with the
chosen range of values for important RF model parameters}
\label{grid}
\begin{tabular}{ll}
\hline
Model Parameter & Chosen Values to Select From \\
\hline

$bootstrap$ & True, False\\
(Whether bootstrapping the samples)
& \\
$ccp\_alpha$ & $2^{-10}$, $2^{-9}$, $2^{-8}$, $2^{-7}$, $2^{-6}$, $2^{-5}$, $2^{-4}$, $2^{-3}$, $2^{-2}$, $2^{-1}$, 0\\
(Complexity parameter)
&  \\
$max\_depth$ &10, 20, 30, 40, 50, 60,70, 
  80, 90, 100, 110, None\\
(Maximum depth of the tree) & \\
$max\_features$ &`auto', `sqrt'\\
(Number of features in each tree) & \\
$min\_samples\_leaf$ & 1, 2, 4 \\
(Minimum samples required for a leaf node) &\\
$min\_samples\_split$ & 2, 5, 10\\
(Minimum samples required to split a node) & \\
$n\_estimators$ &100, 200, 300, 400, 500, 600, 700, 800, 900, 1000\\
(Number of decision trees) & \\
\hline
\end{tabular}
\end{table}

For each cluster we chose the RF model parameters in a way such that the model has the maximum precision. We first divided the training data for each cluster in a test and train subset in a ratio of $30:70$. Then we used the train subset to fine-tune the parameters of the RF model using cross validation. We made a grid with the possible range of values for important model parameters (i.e. number of trees in RF, maximum depth of a tree, minimum samples needed for a split, minimum sample for a leaf node etc), which are reported in Table \ref{grid}. Then we applied a randomized search 5-fold cross validation in the train subset with 100 iteration which in total builds 500 models with randomly chosen parameters from the grid and select the model which resulted in maximum precision.  Finally, we estimated the precision of our selected model using the test subset, which is shown in Table \ref{memdata}.

\subsection{Prediction}

After extracting the Gaia DR2 data within the search radius as specified in Table \ref{memdata}, we subtracted the identified members as we already have them in the training data. In our training data for each cluster we defined membership probability as  1 for the member stars and 0 for the non-member stars. We  then ran the chosen RF classification model on the data to classify them as either members or non-members. We also assigned a membership probability for the stars by measuring the percentage of trees that classify them as members. 

\begin{table}[h!]
\small
\caption{Prediction from the Random Forest Model}
\label{memdata}
\begin{tabular}{lcccccccc}
\hline

 &    & Members   & Members &  Non-Member  & Search & New   & Precision & Ratio of \\
 Cluster & Radius &  before filter  & after filter &  radius &  radius &   Members  &  & new to CG \\
     & deg &     &     &   deg   & deg  &   & \%    &   \\
\hline
NGC 581 & 0.17 & 306 & 290 &  0.7-0.8 & 0.34 &  525 & 86 & 1.81\\
NGC 1893& 0.41 & 494 & 218 & 1.0-1.1  & 0.82 & 774  & 93 & 3.55\\
NGC 2244&0.67 & 1701 &1192 &  1.4-1.5& 1.33 &3043 & 88 & 2.55\\
NGC 2264 &0.19&186   &179  & 1.0-1.1   &0.60 & 514 & 99 & 2.87\\
NGC 3293 & 0.20 & 657    & 617  & 0.7-0.8 & 0.40 & 1089 & 94  & 1.76\\
NGC 6231&0.47&1580   &1354 & 0.95-1.0  &0.94 &2710 & 92  & 2.00\\
NGC 6823&0.2&236     &220  & 0.7-0.8 &0.40  &304  & 93  & 1.38\\
NGC 6913&0.3&170    &170 &  0.7-0.8 & 0.60 & 536 & 95 & 3.15\\
IC 1805 & 0.33 & 456 & 430 &  0.7-0.8 & 0.66 & 1104 & 90 & 2.57\\
\hline
\end{tabular}
\end{table}


\section{Results and Discussions}
Using the RF method we have almost doubled the number of members in our sample of nine clusters as shown in Table \ref{memdata}.  The Figures \ref{pmall} to \ref{sd1}  show the plots obtained from our analysis. 

Figures \ref{pmall}  and \ref{parall} show the proper motion and parallax  distribution for the CG members (orange) and the new members (blue). We can see that the new members are distributed very closely to the CG (training) members in all of these 3 parameters ($\mu_{\alpha}, \mu_{\delta}, \omega$). 
Figure \ref{sky} shows the distribution of the CG members and  new members in sky coordinates. We can see that there are many new members found in the outer regions of the clusters.
Figure \ref{cmall} shows  the CMDs obtained for all the nine clusters with the CG members in orange and the new members in blue. Though we did not use any photometric information while building the model and predicting new stars, all new members agree with the CMD of CG members for all nine clusters, which further validates the accuracy of our member classification. Moreover, as we did not use the photometric information while identifying new members, we can find non-main sequence stars in our sample of clusters from the CMDs. 

In order to visualize the substructures, instead of making a radial profile, we made a kernel density estimation (KDE) contour plot for both ($\alpha,\delta$) space and ($\mu_{\alpha},\mu_{\delta}$) space. 
The color bars in these figures represent the probability density, which is the normalized density of the stars for a given range of ($\alpha, \delta$) or ($\mu_{\alpha}, \mu_{\delta}$). We can get the probability (i.e. frequency) of finding stars within a given range of ($\alpha, \delta$) or ($\mu_{\alpha}, \mu_{\delta}$) by taking the product of the area with their probability density. Integration of the probability density over all the space  will be exactly 1. If we multiply the color bar with the total number of stars for a given plot, we will get the density of stars (count/sq.deg).

Figure  \ref{skyall1} shows the 4 clusters for which we found additional clump(s) besides the cluster center. For all other clusters, the combined KDE plots show a Gaussian distribution as expected (Fig. \ref{skyall2}).  Some of this sub-structure were evident in the training members, but  got further enhanced in the combined sample with the addition of new members. As the clumps are overlapped with each other, it not not very straightforward to find the size or radius for each clump. So we found the center of the clusters and the half radius $R50$, defined as the distance from the center within which, 50\% of the member stars lie. We reported the center and R50 of the central clump for GC members and combined members along with the center of the additional clumps for combined members in Table \ref{cdata}. These clumps show the substructures and fractions in the cluster and they can be further studied to understand the mass segregation across the clusters. 

Similarly, in proper motion space, we also found some additional clumps as seen in Fig. \ref{pmall1}. The presence of more than one clump in the proper motion space indicates the effects of dynamic interactions inside a open star cluster. These clumps can be further studied to understand the underlying dynamics of clusters. Figure \ref{pmall2} shows the KDE plot in proper motion space for the remaining clusters where we see the expected Gaussian distributions.  

The stellar distribution of the stars in the clusters is shown in Fig. \ref{sd1}. We can see that many clusters show a bimodal distribution for the combined members, some of them can be due to the presence of clumps outside the central region. For example, For example, for NGC 2264, we can see that the second peak is near 0.4~deg, which is consistent with the finding that the lower clump center is around 0.4~deg from the central clump as shown in Fig. \ref{skyall1} and reported in Table \ref{cdata}. But we can also see bimodal distribution for the clusters where we did not find any major clump (i.e. NGC 581, NGC 6231), which might be due to a possible asymmetry in the stellar distribution in the clusters.
\begin{center}
\begin{table}[h!]
\small
\caption{Cluster and sub-clump centers and half radius, $R50$. $R50$ is defined as the radius within which 50\% of the members are situated. For each cluster, the first row corresponds to the central clump and the following rows corresponds to the additional subclumps.}
\label{cdata}
\begin{tabular}{l|ll|ll}
\hline
Cluster &CG Members  & & Combined members& \\ 
& Center ($\alpha$, $\delta$) & $R50$ &Center ($\alpha$, $\delta$)& $R50$ \\ 
\hline
NGC 1893 & (80.72, 33.44) & 0.17 &(80.75, 33.47) &0.61\\
         &              &      &(80.86, 33.96) & \\
\hline
NGC 2244 &(98.02, 4.90)   & 0.37 &(98.03, 4.90)  &0.90\\
         &              &      &(98.74, 5.40)  &\\
         &              &      &(98.37, 5.55)  &\\
\hline
NGC 2264 &(100.26, 9.88)  &0.08  & (100.27, 9.49) &0.33\\
         &              &      & (100.23, 9.87) &\\
\hline
NGC 6913 &(305.95, 38.50) &0.17  &(305.94, 38.50)&0.45\\
&&&(305.39, 39.60)&\\
\hline
NGC 581 &(23.36, 60.66)&0.11&(23.35, 60.65)&0.27\\
\hline
NGC 3293 &(158.96, -58.24)&0.11& (158.96, -58.24)&0.34\\
\hline
NGC 6231 &(253.56, -41.83) &0.24& (253.56, -41.82)&0.66\\
\hline
NGC 6823 &(295.79, 23.29) &0.10&(295.79, 23.30)&0.24\\
\hline
IC 1805&(38.22, 61.47)&0.27&(38.22, 61.51)&0.60\\
\hline
\end{tabular}
\end{table}
\end{center}

As seen in Table \ref{memdata}, the number of new members found are more than three times the number (after filtering) of CG members for NGC 1893 and NGC 6913, while for NGC 6823 the ratio was less than 1.5. 

\begin{figure}
    \centering
    \includegraphics[width=0.8\linewidth]{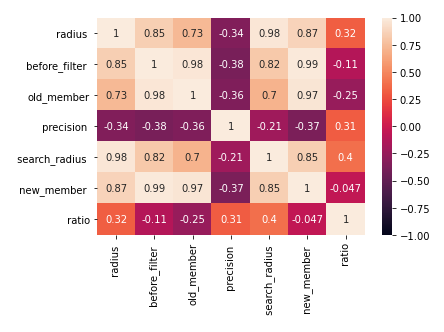}
    \caption{Correlations between parameters in the study}
    \label{corr}
\end{figure}

For our sample of nine clusters, we found that the precision appears to be mildly anti-correlated to the search radius, number of stars used to train the model and the number of new members (with the outlier being NGC 581). Precision appears to be correlated to the ratio of the new members to the CG members. 
For the  ratio of new to CG members there appears to be a anti-correlation to the number of stars used to train the model and a positive correlation to the search radius.  
The correlation between stars used to train the model  and the number of new members is very tight and has a value of 0.97. The detailed correlation values are described in Fig. \ref{corr}.

As our sample size is small, these correlations are mere indicators and are subject to change with larger samples\footnote{The combined membership data for all the nine clusters can be requested from PH.}.

\section{Conclusion}
We have used RF to estimate membership of a larger sample of stars in Gaia DR2 data for nine open clusters. This kind of analysis has the following major advantages:
\begin{itemize}
    \item Our results indicate that this machine-learning-based method is highly suitable for membership determination of open clusters in high dimensional feature space.     
    \item The sample of stars in clusters can be increased by a large factor, almost 2--3 times. This improves our accuracy in determining various parameters of a star cluster ranging from distance, extinction and mass function. The sizes of the studied clusters also increased with the increase in membership and we can study the outer regions of  clusters. The method may find not only likely cluster members but also likely escaped members, which may lie outside the tidal radius of the cluster. 
    \item
    This method makes it possible to find sub-structure in position space as well as velocity  space of the cluster as the members lie in regions beyond the reported sizes of clusters and hence we can identify sub-clumps as in NGC 2244, NGC 2264 and NGC 6913.
    \item
    As we have not used photometric data while estimating the membership, we can identify variables, pre-main sequence stars (in NGC~1893, NGC~3293), unresolved binary sequences (in NGC~6231) as well as all other possible non main-sequence members of the cluster from its CMD.
    
    This work strongly indicates that the RF method has very good potential applications in star cluster studies. 
    
\end{itemize}

\section{Acknowledgments}
This work has made use of data from the European Space Agency (ESA) mission {\it{Gaia}} (\url{https://www.cosmos.esa.int/gaia}), processed by the {\it{Gaia}} Data Processing and Analysis Consortium (DPAC, https://www.cosmos.esa.int/ web/gaia/dpac/consortium). Funding for the DPAC institutions, in particular the institutions participating in the {\it{Gaia}} Multilateral Agreement. This work has made use of data from the European Space Agency (ESA) mission {\it{Gaia}} (\url{https://www.cosmos.esa.int/gaia}), processed by the {\it{Gaia}} Data Processing and Analysis Consortium (DPAC,  {https: //www.cosmos.esa.int/web/gaia/dpac/consortium}). Funding for the DPAC institutions, in particular the institutions participating in the {\it{Gaia}} Multilateral Agreement. 
Virtual observatory tools like Aladin, Vizier and TOPCAT have been used in the analysis.
Astropy, Scikit Learn, Seaborn, Numpy and Pandas packages in python have been used in the analysis and visualizations.
MM is very thankful to Dr. Rohan Sekhar, Associate Professor of Computational Sciences in Minerva Schools at KGI for providing necessary support and insights on the RF model.


\bibliographystyle{spphys}

\bibliography{biblio}

\end{document}